\documentclass[prd,aps,nofootinbib,floatfix,10pt]{revtex4}
\usepackage{amsmath,graphicx,epsfig,amssymb}
\usepackage{inputenc}
\usepackage[usenames]{color}
\usepackage{ulem} %% for strike-through
\usepackage{bigstrut}
\usepackage{multirow}
\usepackage{subfigure}

\allowdisplaybreaks

\begin{document}
	
\title{SU(3) symmetry and its breaking effects in semileptonic heavy baryon decays}
\author{Xiao-Gang He$^{1,2,3,4}$,
Fei Huang$^{2}$,
Wei Wang$^{2}$, and Zhi-Peng Xing$^{1}$  }
\affiliation{$^{1}$  Tsung-Dao Lee Institute, Shanghai Jiao Tong University, Shanghai 200240, China}
\affiliation{$^{2}$
                   INPAC, Key Laboratory for Particle Astrophysics and Cosmology (MOE), Shanghai Key Laboratory for Particle Physics and Cosmology,
                   School of Physics and Astronomy, Shanghai Jiao Tong University, Shanghai
		  200240, China
		 }		
\affiliation{$^{3}$
		   Physics Division,National Center for Theoretical Sciences, Taipei 10617, Taiwan }		
\affiliation{$^{4}$
		    Department of Physics, National Taiwan University, Taipei 10617, Taiwan }

\begin{abstract}
We employ the flavor SU(3) symmetry to analyze semileptonic decays of anti-triplet charmed baryons ($\Lambda^{+}_{c},\Xi_{c}^{+,0}$) and find that the experimental data on ${\cal B}(\Lambda_c\to\Lambda \ell^+\nu_{\ell})$ implies $\mathcal{B}(\Xi_c^0\to\Xi^-e^+\nu_{e})=(4.10\pm0.46)\%$ and $\mathcal{B}(\Xi_c^0\to\Xi^-\mu^+\nu_{\mu})= (3.98\pm0.57)\%$. When this prediction is confronted with recent experimental results from Belle collaboration $\mathcal{B}(\Xi_c^0\to\Xi^-e^+\nu_e)=(1.31\pm0. 04\pm0. 07\pm0. 38)\%$ and $\mathcal{B}(\Xi_c^0\to\Xi^-\mu^+\nu_{\mu})= (1.27\pm 0.06\pm0.10\pm 0.37)\%$, it is found that the SU(3) symmetry is severely broken. We then consider the generic SU(3) breaking effects both in decay amplitudes and the mixing effect between the anti-triplet and sextet charmed baryons. We find that the pertinent data can be accommodated in different scenarios but the breaking effects are inevitably large. In some interesting scenarios, we also explore the testable implications in these scenarios which can be stringently tested with more data become available. Similar analyses are carried out for the semileptonic decays of anti-triplet beauty baryons to octet baryons and anti-triplet charmed baryons. The validity of SU(3) for these decays can also be examined when data become available.
\end{abstract}
\maketitle
	
%\tableofcontents
%%%%%%%%%%%%%%%%%%%%%%%%%%%%
%%%%%%%%%%%%%
%%%%%%%%%%%%%%%%%%%%%%%%%%%%%%%%%%%%%%%%
\section{introduction}
Weak decays of heavy baryons carrying a charm quark have been studied extensively on both experimental and theoretical aspects~\cite{BESIII:2020nme},  as they supply a  platform for the study of strong and weak interactions in the standard model (SM).
On the experimental side, data on charmed baryons decays from BESIII~\cite{Ablikim:2015prg,Ablikim:2016vqd},  Belle \cite{Belle:2021crz} and ALICE~\cite{ALICE:2021bli} collaborations  provided important information to extract the CKM matrix element.  Belle collaboration has provided a measurement of the $\Xi^0_{c}$ branching fractions very recently~\cite{Belle:2021crz}:
\begin{eqnarray}
\mathcal{B}_{\rm Belle}(\Xi^{0}_{c}\rightarrow\Xi^{-} e^{+}\nu_{e})=(1.31\pm0.04\pm0.07\pm0.38)\% \; ,\nonumber \\
\mathcal{B}_{\rm Belle}(\Xi^{0}_{c}\rightarrow\Xi^{-} \mu^{+}\nu_{\mu})=(1.27\pm0.06\pm0.10\pm0.37)\%\;,\label{eq:Belle-data}
\end{eqnarray}
which is about a factor of 2 more precise than the ALICE result:
\begin{eqnarray}
\mathcal{B}_{\rm  ALICE}(\Xi^{0}_{c}\rightarrow\Xi^{-} e^{+}\nu_{e})=(2.5\pm0.8)\%\;.
\end{eqnarray}
This comes from the ALICE  measurement of $\mathcal{B}(\Xi^{0}_{c}\rightarrow\Xi^{-} e^{+}\nu_{e})/\mathcal{B}(\Xi^{0}_{c}\rightarrow\Xi^{-}\pi^{+})=1.38\pm0.14\pm0.22$~\cite{ALICE:2021bli} and Belle data $\mathcal{B}(\Xi^{0}_{c}\rightarrow\Xi^{-}\pi^{+})=1.8\pm 0.7\%$~\cite{Belle:2018kzz}. We anticipate the difference between the above results can be clarified with the improvement of the experimental accuracy and the promising prospects on charmed baryons in the future. The available data on the decays from the anti-triplet heavy baryons to the octet baryons have been collected in Table~\ref{exp}, while the branching fraction $\mathcal{B}(\Xi^{0}_{c}\rightarrow\Xi^{-} e^{+}\nu_{e})=(1.54\pm0.35)\%$ listed is obtained by averaging the Belle and ALICE data.

On the theoretical side, one can apply the SU(3) flavor symmetry to analyze the semileptonic decays and obtain some model-independent relations among different decays~\cite{Gronau:1995hm,Li:2012cfa,Cheng:2014rfa,Zhou:2016jkv,Muller:2015lua,Geng:2017mxn,Geng:2018plk,Geng:2019bfz,He:2015fwa,Lu:2016ogy,Geng:2018rse,Gronau:2013mza,Savage:1989qr}.  For semileptonic charmed baryon decays, we have
\begin{eqnarray}
\Gamma(\Xi^{0}_{c}\to \Xi^{-}\ell^+\nu_{\ell})=\Gamma(\Xi^{+}_{c}\to \Xi^{0}\ell^+\nu_{\ell})=\frac{3}{2}\Gamma(\Lambda^{+}_{c}\to \Lambda^{0}\ell^+\nu_{\ell})\;.\label{su3analysis}
\end{eqnarray}
Since the irreducible amplitude can be extracted by fitting data, the SU(3) analysis bridges experimental data and the dynamical approaches like Lattice QCD~\cite{Meinel:2016dqj,Meinel:2017ggx,Zhang:2021oja,Detmold:2015aaa} and model-dependent calculations~\cite{Aliev:2021wat,Zhao:2018zcb,Liu:2010bh,Azizi:2011mw,Lu:2000em,Cheng:1991sn,Faustov:2016yza,Li:2016qai,Guo:2005qa}.
We adopt the experiment data on $\Lambda_{c}^{+}$ semileptonic decays and the SU(3) relations with the lifetimes $\tau_{\Lambda_{c}^{+}}=2.024\times 10^{-13}$s, $\tau_{\Xi_{c}^{0}}=1.53\times 10^{-13}$s, $\tau_{\Xi_{c}^{+}}=4.56\times 10^{-13}$s~\cite{Zyla:2020zbs}. Then we obtain the branching ratios of $\Xi_{c}^{0,+}$ shown in Table~\ref{exp}, from which one can find an obvious deviation between experiments and theory.

 %%%%%%%%%%%%%%%%%%%
\begin{table}[htbp!]\label{exp}
\caption{Experimental data and SU(3) symmetry analysis of anti-triplet charmed baryon decays. The SU(3) predictions for ${\cal B}(\Xi_{c}\to \Xi \ell^+\nu_{\ell})$ are obtained by fitting the first two experimental data. }
\begin{tabular}{|c|c|c|c|c|c|c|c|c|c|c|}\hline\hline
\multirow{2}{*}{channel} &\multicolumn{2}{c|}{ branching ratio$(\%)$} \cr\cline{2-3}
&experimental data& SU(3) symmetry\\\hline
$\Lambda^{+}_{c}\to \Lambda^{0}e^+\nu_{e} $ & $3.6\pm0.4$~\cite{Zyla:2020zbs}&$3.6\pm0.4$~(input)\\\hline
$\Lambda^{+}_{c}\to \Lambda^{0}\mu^+\nu_{\mu} $ & $3.5\pm0.5$~\cite{Zyla:2020zbs}&$3.5\pm0.5$~(input)\\\hline
$\Xi^{+}_{c}\to \Xi^{0}e^+\nu_{e} $ & $2.3\pm1.5$~\cite{Zyla:2020zbs}& $12.17\pm 1.35$\\\hline
$\Xi^{0}_{c}\to \Xi^{-}e^+\nu_{e} $ & $ 1.54\pm0.35$~\cite{Belle:2021crz,ALICE:2021bli}& $4.10\pm 0.46$\\\hline
$\Xi^{0}_{c}\to \Xi^{-}\mu^+\nu_{\mu} $ & $ 1.27\pm0.44$~\cite{Belle:2021crz} & $3.98\pm0.57$\\\hline
\end{tabular}
\end{table}
 %%%%%%%%%%%%%%%\cite{Belle:2021crz} and ALICE~\cite{ALICE:2021bli} %%%%

It should be noted that the flavor SU(3) symmetry is an approximate symmetry, since u, d, and s quarks have different masses which breaks SU(3) symmetry~\cite{Geng:2018bow}. For a more accurate analysis,   SU(3) breaking effects should be included, which is the main focus of this work.  Compared to the strange quark mass $m_{s}$, the up and down quark masses $m_{u,d}$ are much smaller and thus can be neglected. Therefore the s quark mass is the major source for flavor SU(3) symmetry breaking. In this work, we carry out an analysis with the leading-order SU(3)  breaking effects on semileptonic anti-triplet charmed baryons decays and explore the scenarios in which recent experimental measurements can be consistently accommodated.

The rest of this paper is organized as follows. In Sec.~II, we give the theoretical framework for SU(3) symmetry and study symmetry breaking in semileptonic decays of anti-triplet heavy baryons for the process of $c\rightarrow d/s$. In Sec.~III, we also obtain numerical results using the SU(3) symmetry term and analyze the SU(3) symmetry breaking effect for the process of $b\rightarrow c/u$. A brief conclusion will be presented in the last section.
\section{SU(3) symmetry for semileptonic anti-triplet charmed baryon decays}

\begin{figure}[htbp!]
  \begin{minipage}[t]{0.6\linewidth}
  \centering
  \includegraphics[width=1.0\columnwidth]{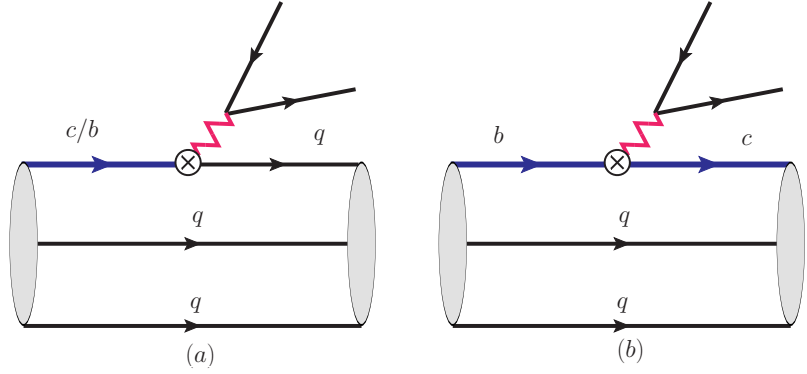}
    \end{minipage}
\caption{The Feynman diagram of anti-triplet heavy baryons induced by $c\to d/s$, $b\to u$(left), and $b\to c$(right).}
\label{feynman}
\end{figure}

In the flavor SU(3) symmetry limit,  hadron multiplets can be classified according to the SU(3) irreducible representation. Baryons with a charm quark and two light quarks can have  $3\otimes3=\bar{3}\oplus6$ representations. The anti-triplet $\bar 3$ semileptonic baryon  $(\Lambda_c^+,\Xi_c^+,\Xi_c^0)$  decays are our focus here, whose quark level Feynman diagrams are shown in Fig~\ref{feynman}(a).
In the SM the low-energy effective Hamiltonian for these decays is given as
\begin{eqnarray}
{\cal H}_{c\to d/s}&=&\frac{G_F}{\sqrt2} \left[V_{cq}^* \bar q  \gamma^\mu(1-\gamma_5)c ~\bar \nu_{\ell}\gamma_\mu(1-\gamma_5)\ell\right] +h.c.,
\end{eqnarray}
where $q=d,s$ and $G_F$ is the Fermi-constant. $V_{cq}$ is CKM matrix element. With the help of helicity amplitude method~\cite{Huang:2021ots}, the decay transition amplitude can be written as
\begin{eqnarray}
{\cal A}(B_c\to B_q\ell^+\nu_{\ell})&=&\frac{G_F}{\sqrt{2}}V_{cq}^{*}\langle B_q| \bar{q}\gamma^\mu(1-\gamma_5)c|B_c\rangle \langle \ell^{+}\nu_{\ell}|\bar{\nu}_{\ell}\gamma^\nu(1-\gamma_5)\ell|0 \rangle g_{\mu\nu},
\end{eqnarray}
with the decomposition of $g_{\mu\nu}$,
\begin{align}
g_{\mu\nu}=&-\sum_{\lambda=0,\pm1}\epsilon_{\mu}^*(\lambda)\epsilon_{\nu}(\lambda)+\epsilon_{\mu}^*(t)\epsilon_{\nu}(t)\;,\notag\\
\epsilon_{\mu}(t)=&\frac{q^\mu}{\sqrt{q^2}},
\end{align}
where the $\epsilon_{\mu}(\lambda)$ is transverse($\lambda=\pm1$) or longitudinal($\lambda=0$) polarization states and $\epsilon_{\mu}(t)$ is timelike polarization states.

The above amplitude can be decomposed  into the Lorentz invariant hadronic and leptonic matrix elements:
 \begin{align}
	{\cal A}(B_c\to B_q\ell^+\nu)=& \frac{G_F}{\sqrt{2}}V_{cq}^{*}\langle B_q| \bar{q}\gamma^\mu(1-\gamma_5)c|B_c\rangle \langle \ell^{+}\nu_{\ell}|\bar{\nu_{\ell}}\gamma^\nu(1-\gamma_5)\ell|0 \rangle g_{\mu\nu} \nonumber\\
	=& \frac{G_F}{\sqrt{2}}V_{cq}^{*} \left( -\sum_{\lambda_w=0,\pm1}H_{\lambda,\lambda_w} L_{\lambda_w}+H_{\lambda,t} L_{t}\right),\\ \notag
H_{\lambda,\lambda_w}=&\langle B_q|\bar q  \gamma^\mu(1-\gamma_5)c|B_c\rangle \epsilon^*_\mu(\lambda_w)\;,\\ \notag
L_{\lambda_w}=&\langle \ell^{+}\nu_{\ell}|\bar{\nu_{\ell}}\gamma^\nu(1-\gamma_5)\ell|0 \rangle \epsilon_{\nu}(\lambda_w)\;,
\end{align}
where $H_{\lambda,\lambda_w}$($L_{\lambda_w}$) is hadronic(leptonic) helicity amplitude, $\lambda_{(W)}$($0,\pm1,t$) corresponds to the helicity of the daughter baryon ($W$) and the $\epsilon_\mu(\lambda_W)$ is the polarization vector of $W$ boson.

%%%%%%%%%%%%%%%%%%%%
%%%%%%%%%%%%%%%%%%%%

In the SM, charmed baryons can decay into octet baryons. The SU(3) anti-triplet and octet matrix are denoted by
\begin{eqnarray}\
T_{c\bar3}=
\begin{pmatrix}
0& \Lambda_c^+ &\Xi_c^+ \\
-\Lambda_c^+ & 0&\Xi_c^0\\
-\Xi_c^+& -\Xi_c^0&0
\end{pmatrix},\quad T_8&=&
\begin{pmatrix}
\frac{\Sigma^0}{\sqrt{2}}+\frac{\Lambda^0}{\sqrt{6}}& \Sigma^+ &p \\
\Sigma^- & -\frac{\Sigma^0}{\sqrt{2}}+\frac{\Lambda^0}{\sqrt{6}}&n\\
\Xi^-& \Xi^0&-\frac{2\Lambda^0}{\sqrt{6}}
\end{pmatrix}.
\end{eqnarray}
Tree operators of charm quark semileptonic decays into light quarks are categorized into $c \rightarrow d/s$.
Therefore under the flavor SU(3) symmetry, the low-energy effective Hamiltonian can be decomposed in terms of $H_{3}$ shown as:
\begin{eqnarray}
(H_3)^1=0,\quad(H_3)^2=V^*_{cd},\quad(H_3)^3=V^*_{cs}.
\end{eqnarray}
The corresponding helicity amplitude can be written as:
\begin{eqnarray}
H_{\lambda,\lambda_w}=a_1^{\lambda,\lambda_w} \times(T_{c\bar3})^{[ij]}(H_3)^k \epsilon_{ikm}(T_8)^m_j, \label{eq:Bc3toB8}
\end{eqnarray}
where the $a_1^{\lambda,\lambda_w}$ represents SU(3) irreducible nonperturbative amplitude. The $a_1^{\lambda,\lambda_w}$ can be expressed by the form factors
\begin{eqnarray}
a_1^{\lambda,\lambda_w}&=&\bar u(\lambda)\bigg[f_1\gamma^\mu+f_2\frac{i\sigma^{\nu\mu}}{M_i}q^\nu+f_3\frac{q^\mu}{M_i}\bigg]u(\lambda_i)\epsilon^*_\mu(\lambda_w) \notag\\
&&-\bar u(\lambda)\bigg[f^\prime_1\gamma^\mu+f^\prime_2\frac{i\sigma^{\nu\mu}}{M_i}q^\nu+f^\prime_3\frac{q^\mu}{M_i}\bigg]\gamma_5u(\lambda_i)\epsilon^*_\mu(\lambda_w).  \label{hilicity}
\end{eqnarray}
Here $ u(\lambda_i)$ is the spinor of charmed baryons, $u(\lambda) $ is the spinor of the final state, and $f_{i}(i=1,2,3)$,  $f^{\prime}_{i}(i=1,2,3)$ are the form factors. In the heavy quark limit~\cite{Manohar:2000dt},  the $f_2$, $f_3$, $f_2^\prime$, $f_3^\prime$ are suppressed by $1/m_{B_c}$, and only one independent  form factor exists if the large-recoil symmetry is adopted~\cite{Mannel:2011xg,Wang:2011uv}. Actually, a previous calculation~\cite{Zhao:2018zcb} also indicates that  the form factor of vector parameter $f_1$ and axial-vector parameter $f^\prime_1$ have dominant contributions to the heavy baryon decay processes.
Thus we neglect $f_2$, $f_2^\prime$, $f_3$, and $f_3^\prime$ in our later calculations.
Expanding Eq.~\eqref{eq:Bc3toB8}, one obtains the relations between the helicity amplitudes of different channels of anti-triplet charmed baryons, which are presented in Table~\ref{SU3}.
%%%%%%%%%%%%%%%%%%%
 \begin{table}[htbp!]\label{SU3}
 \caption{Decay amplitudes charmed baryons $\Xi_{c}$ and $\Lambda_{c}$ decays into an octet baryons. These amplitudes can be obtained by expanding Eq.~\eqref{eq:Bc3toB8}.  }
\begin{tabular}{|c|c|c|c|c|c|c|c}\hline\hline
channel & amplitude \\\hline
$\Lambda^{+}_{c}\to \Lambda^{0}\ell^+\nu_{\ell} $ & $ -\sqrt{\frac{2}{3}} a_1^{\lambda,\lambda_w} V_{\text{cs}}^{*}$\\\hline
$\Lambda^{+}_{c}\to n\ell^+\nu_{\ell}  $ & $ a_1^{\lambda,\lambda_w} V_{\text{cd}}^{*}$\\\hline
$\Xi^{+}_{c}\to \Sigma^{0}\ell^+\nu_{\ell}  $ & $ \frac{a_1^{\lambda,\lambda_w} V_{\text{cd}}^{*}}{\sqrt{2}}$\\\hline
$\Xi^{+}_{c}\to \Lambda^{0}\ell^+\nu_{\ell}  $ & $ -\frac{a_1^{\lambda,\lambda_w} V_{\text{cd}}^{*}}{\sqrt{6}}$\\\hline
$\Xi^{+}_{c}\to \Xi^{0}\ell^+\nu_{\ell}  $ & $ -a_1^{\lambda,\lambda_w} V_{\text{cs}}^{*}$\\\hline
$\Xi^{0}_{c}\to \Sigma^{-}\ell^+\nu_{\ell}  $ & $ a_1^{\lambda,\lambda_w} V_{\text{cd}}^{*}$\\\hline
$\Xi^{0}_{c}\to \Xi^{-}\ell^+\nu_{\ell}  $ & $ a_1^{\lambda,\lambda_w} V_{\text{cs}}^{*}$\\\hline
\end{tabular}\
\end{table}
%%%%%%%%%%%%%%%%%{eq:Bc3toB8}
In the SU(3) symmetry limit, the branching fractions of $\Xi^{+}_{c}\to \Xi^{0}\ell^+\nu_{\ell}  $ and $\Xi^{0}_{c}\to \Xi^{-}\ell^+\nu_{\ell}  $ can be predicted by using experimental data of $\Lambda^{+}_{c}\to \Lambda^{0}\ell^+\nu_{\ell}  $ which are given in Table~\ref{exp}. 
To shed further light on the decay dynamics, we take the pole model as an illustration to access the $q^2$ dependence of form factors~\cite{Zhao:2018mrg}
\begin{eqnarray}
f_i(q^2)=\frac{f_i}{1-\frac{q^2}{m_p^2}},\label{eq:pole}
\end{eqnarray}
where $f_i=f_i(q^2=0)$ and $m_p=2.061{\rm GeV}$, which is the average mass of $D$ and $D_s$.
The differential decay widths can be expressed by these form factors,
\begin{eqnarray}
\frac{d\Gamma}{dq^{2}}  &=& \frac{(m_{\ell}^2-q^2)^2\sqrt{\lambda}G_F^2 V_{SU(3)}^2}{284\pi^3 M^3(q^2)^3}\bigg[(f_1(q^2))^2 \times (3 s_+ m_{\ell}^2 \left(q^2+s_-\right)+s_- \left(3 q^2+s_+\right) \left(m_{\ell}^2+2 q^2\right))\nonumber\\
&& + (f_1^{\prime }(q^2))^2 \times  (3 s_- m_{\ell}^2 \left(q^2+s_+\right)+s_+ \left(3 q^2+s_-\right) \left(m_{\ell}^2+2 q^2\right))\bigg].\label{Gamma}
\end{eqnarray}
Here $s_-=(M-M^\prime)^2-q^2, s_+=(M+M^\prime)^2-q^2,$ and $\sqrt{\lambda}=\sqrt{s_-s_+}$.  $M$ and $M^\prime$ are the mass of $B_c$ and $B_q$, respectively. $m_l$ is the lepton mass. $V_{SU(3)}$ is the SU(3) factor coming from the coefficient of $a_1^{\lambda,\lambda_w}$ in Table~\ref{SU3}. For instance in the first process in Table ~\ref{SU3}, $V_{SU(3)}=-\sqrt{2/3}V_{cs}^{*}$.

Using the amplitudes, we can fit the parameters $f_1$ and $f^\prime_1$ with experimental data. In the fit, we use the experimental values for the particle masses. The fitted results are shown in Table~\ref{data}.
Obviously, the $\chi^2$ in fitting is too large to be considered as a good fit, which implies that the SU(3) symmetry is not a good symmetry for charmed baryon decays.
%%%%%%%%%%%%%%%%%%%%%%
%%%%%%%%%%%%%%%%%%%%%%
\begin{table}[htbp!]
\caption{Experimental and fit data of anti-triplet charmed baryons decays.}\label{data}\label{data}
\begin{tabular}{|c|c|c|c|c|c|c|c|c|c|c|}\hline\hline
\multirow{2}{*}{channel} &\multicolumn{2}{c|}{ branching ratio$(\%)$} \cr\cline{2-3}
&experimental data& fit data\\\hline
$\Lambda^{+}_{c}\to \Lambda^{0}e^+\nu_{e} $ & $3.60\pm0.40$&$1.94\pm0.18$\\\hline
$\Lambda^{+}_{c}\to \Lambda^{0}\mu^+\nu_{\mu} $ & $3.5\pm0.5$&$1.87\pm0.176$\\\hline
$\Xi^{+}_{c}\to \Xi^{0}e^+\nu_{e} $ & $2.3\pm1.5$&$6.53\pm0.60$\\\hline
$\Xi^{0}_{c}\to \Xi^{-}e^+\nu_{e} $ & $ 1.54\pm0.35$&$2.17\pm0.20$\\\hline
$\Xi^{0}_{c}\to \Xi^{-}\mu^+\nu_{\mu} $ & $ 1.27\pm0.44$&$2.09\pm0.19$\\\hline
$\chi^2/d.o.f=14.3$&$f_1=1.05\pm0.30$&$f^\prime_1=0.11\pm0.95$\\\hline
\end{tabular}
\end{table}
%%%%%%%%%%%%%%%%%%%%%%
%%%%%%%%%%%%%%%%%%%%%%
In the previous fit, we have neglected the possible SU(3) breaking effects. Because the light u, d, and s quarks have different masses, the SU(3) symmetry is broken. Neglecting the masses of u and d quark the mass matrix $M$ can be written as:
\begin{eqnarray}
M=\begin{pmatrix}
m_u& 0 &0\\
0& m_d&0\\
0& 0&m_s
\end{pmatrix}\sim m_s
\begin{pmatrix}
0& 0 &0\\
0& 0&0\\
0& 0&1
\end{pmatrix}
=m_s\times\omega. \;\label{SU3SBT}
\end{eqnarray}
We can obtain the modified helicity amplitude as
\begin{eqnarray}
H_{\lambda,\lambda_W}=&&a_1^{\lambda,\lambda_w} \times(T_{c\bar3})^{[ij]}(H_3)^k \epsilon_{ikm}(T_8)^m_j+a_2^{\lambda,\lambda_w} \times(T_{c\bar3})^{[in]}(H_3)^k \epsilon_{ikm}(T_8)^m_j \omega_n^j\notag\\&+&a_3^{\lambda,\lambda_w} \times(T_{c\bar3})^{[in]}(H_3)^k \epsilon_{kjm}(T_8)^m_i \omega_n^j+a_4^{\lambda,\lambda_w} \times(T_{c\bar3})^{[in]}(H_3)^k \epsilon_{jim}(T_8)^m_k \omega_n^j\notag\\
&+&a_5^{\lambda,\lambda_w} \times(T_{c\bar3})^{[ij]}(H_3)^k \epsilon_{inm}(T_8)^m_j \omega^n_k.\label{eq:charmsb}
\end{eqnarray}
The $a_1^{\lambda,\lambda_w}$ is SU(3) symmetric irreducible nonperturbative amplitude and $a_2^{\lambda,\lambda_w}$,$\;a_3^{\lambda,\lambda_w}$,$\;a_4^{\lambda,\lambda_w}$,$\;a_5^{\lambda,\lambda_w}$are SU(3) symmetry breaking irreducible nonperturbative amplitudes, which are proportional to $m_s$.  Furthermore, the SU(3) symmetry breaking terms in Eq.~\eqref{eq:charmsb} include the contribution of anti-triplet and sextet charmed heavy baryons mixing terms which correspond to the contribution of $a_{2}^{\lambda,\lambda_w}$,$a_{3}^{\lambda,\lambda_w}$ and $a_{4}^{\lambda,\lambda_w}$.

\subsection{Symmetry breaking in  helicity amplitude}\label{sec11}

The  SU(3) symmetry breaking irreducible nonperturbative amplitudes  $a_2^{\lambda,\lambda_w}$, $\;a_3^{\lambda,\lambda_w}$,$\;a_4^{\lambda,\lambda_w}$, $\;a_5^{\lambda,\lambda_w}$ in Eq.~\eqref{eq:charmsb} can be decomposed in a similar way as that in Eq.~(\ref{hilicity}). Again in our analysis, we only keep the vector and axial vector form factors.

Adding SU(3) symmetry breaking term and expanding the above formula in Eq.~\eqref{eq:charmsb},  one can  obtain the amplitudes of different channels   which are  collected in the `` amplitude I " column of Table~\ref{SU3phy}. It can be seen that the parameters $a_1^{\lambda,\lambda_w}$ and $a_5^{\lambda,\lambda_w}$ always appear together in channels $\Lambda^{+}_{c}\to \Lambda^{0}\ell^+\nu_{\ell} $, $\Xi^{+}_{c}\to \Xi^{0}\ell^+\nu_{\ell}  $ and $\Xi^{0}_{c}\to \Xi^{-}\ell^+\nu_{\ell}  $.
Therefore the SU(3) symmetry breaking irreducible nonperturbative amplitudes $a_2^{\lambda,\lambda_w}$, $a_3^{\lambda,\lambda_w}$, $a_4^{\lambda,\lambda_w}$, and  $a_5^{\lambda,\lambda_w}$ can be parametrized as
\begin{eqnarray}
a_1^{\lambda,\lambda_w}+a_5^{\lambda,\lambda_w}
&=&f_1(q^2)\times\bar u(\lambda)\gamma^\mu u(\lambda_i)\epsilon^*_\mu(\lambda_w)- f^\prime_1(q^2)\times\bar u(\lambda)\gamma^\mu \gamma_5u(\lambda_i)\epsilon^*_\mu(\lambda_w),\notag\\
a_2^{\lambda,\lambda_w}-a_4^{\lambda,\lambda_w}&=&\delta f_1(q^2)\times\bar u(\lambda)\gamma^\mu u(\lambda_i)\epsilon^*_\mu(\lambda_w)-\delta f^\prime_1(q^2)\times\bar u(\lambda)\gamma^\mu \gamma_5u(\lambda_i)\epsilon^*_\mu(\lambda_w),\notag\\
a_3^{\lambda,\lambda_w}&=&\Delta f_1(q^2)\times\bar u(\lambda)\gamma^\mu u(\lambda_i)\epsilon^*_\mu(\lambda_w)-\Delta f^\prime_1(q^2)\times\bar u(\lambda)\gamma^\mu \gamma_5u(\lambda_i)\epsilon^*_\mu(\lambda_w),
\label{hme}
\end{eqnarray}
where the $a_2^{\lambda,\lambda_w}-a_4^{\lambda,\lambda_w}$ is the combination that appears in helicity amplitude $\Xi^{+}_{c}\to \Xi^{0}\ell^+\nu_{\ell}  $ and $\Xi^{0}_{c}\to \Xi^{-}\ell^+\nu_{\ell}$.

\begin{table}[htbp!]
 \caption{Decay amplitudes of charmed baryons anti-triplet decay into an octet baryon.  The amplitudes in column I without $c_1^{\lambda,\lambda_w}$ term come from Eq. (\ref{eq:charmsb}). The effects of $\Xi_c$ and $\Xi_c^\prime$ mixing can be obtained by adding terms proportional to $c_1^{\lambda,\lambda_w}$. }\label{SU3phy}
\begin{tabular}{|c|c|c|c|c|c|c|c}\hline\hline
channel & amplitude I & amplitude II\\\hline
$\Lambda^{+}_{c}\to \Lambda^{0}l^+\nu $ & $ -\sqrt{\frac{2}{3}}( a_1^{\lambda,\lambda_w}+a_5^{\lambda,\lambda_w}) V^*_{\text{cs}}$&$ -\sqrt{\frac{2}{3}}( a_1^{\lambda,\lambda_w}+a_5^{\lambda,\lambda_w}) V^*_{\text{cs}}$\\\hline
$\Lambda^{+}_{c}\to nl^+\nu $ & $ a_1 V^*_{\text{cd}}$& $ a_1 V^*_{\text{cd}}$\\\hline
$\Xi^{+}_{c}\to \Sigma^{0}\ell^+\nu_{\ell}  $ & $\frac{(a_{1}^{\lambda,\lambda_w}+a_{3}^{\lambda,\lambda_w}-a_{4}^{\lambda,\lambda_w}-\frac{c_{1}^{\lambda,\lambda_w}}{\sqrt{2}} \theta) V_{\text{cd}}^{*}}{\sqrt{2}}$& $\frac{(a_{1}^{\lambda,\lambda_w}+a_{3}^{\lambda,\lambda_w}-a_{4}^{\prime\lambda,\lambda_w}) V_{\text{cd}}^{*}}{\sqrt{2}}$\\\hline
$\Xi^{+ }_{c}\to \Lambda^{0}\ell^+\nu_{\ell}  $ & $ -\frac{(a_{1}^{\lambda,\lambda_w}+2a_{2}^{\lambda,\lambda_w}-a_{3}^{\lambda,\lambda_w}-a_{4}^{\lambda,\lambda_w}+\frac{3c_{1}^{\lambda,\lambda_w}}{\sqrt{2}} \theta) V_{\text{cd}}^{*}}{\sqrt{6}}$& $ -\frac{(a_{1}^{\lambda,\lambda_w}+2a_{2}^{\prime\lambda,\lambda_w}-a_{3}^{\lambda,\lambda_w}-a_{4}^{\prime\lambda,\lambda_w}) V_{\text{cd}}^{*}}{\sqrt{6}}$\\\hline
$\Xi^{+ }_{c}\to \Xi^{0}\ell^+\nu_{\ell}  $ & $ -(a_{1}^{\lambda,\lambda_w}+a_{2}^{\lambda,\lambda_w}-a_{4}^{\lambda,\lambda_w}+a_{5}^{\lambda,\lambda_w}+\frac{c_{1}^{\lambda,\lambda_w}}{\sqrt{2}} \theta) V_{\text{cs}}^{*}$& $ -(a_{1}^{\lambda,\lambda_w}+a_{2}^{\prime\lambda,\lambda_w}-a_{4}^{\prime\lambda,\lambda_w}+a_{5}^{\lambda,\lambda_w}) V_{\text{cs}}^{*}$\\\hline
$\Xi^{0 }_{c}\to \Sigma^{-}\ell^+\nu_{\ell}  $ & $ (a_{1}^{\lambda,\lambda_w}+a_{3}^{\lambda,\lambda_w}-a_{4}^{\lambda,\lambda_w}-\frac{c_{1}^{\lambda,\lambda_w}}{\sqrt{2}} \theta) V_{\text{cd}}^{*}$& $ (a_{1}^{\lambda,\lambda_w}+a_{3}^{\lambda,\lambda_w}-a_{4}^{\prime\lambda,\lambda_w}) V_{\text{cd}}^{*}$\\\hline
$\Xi^{0 }_{c}\to \Xi^{-}\ell^+\nu_{\ell}  $ & $ (a_{1}^{\lambda,\lambda_w}+a_{2}^{\lambda,\lambda_w}-a_{4}^{\lambda,\lambda_w}+a_{5}^{\lambda,\lambda_w}+\frac{c_{1}^{\lambda,\lambda_w}}{\sqrt{2}} \theta) V_{\text{cs}}^{*}$& $ (a_{1}^{\lambda,\lambda_w}+a_{2}^{\prime\lambda,\lambda_w}-a_{4}^{\prime\lambda,\lambda_w}+a_{5}^{\lambda,\lambda_w}) V_{\text{cs}}^{*}$\\\hline
\end{tabular}\
\end{table}
%%%%%%%%%%%%%%%%%%%%%%%%%%%%%%%%%%%%%%%%%%%%%%%%%%%%%%%%%%%%%%%%%%%%%%
%%%%%%%%%%%%%%%%%%%%%%%%%%%%%%%%%%%%%%%%%%%%%%%%%%%%%%%%%%%%%%%%%%%%%%%

By using the replacement rule: $f^{(\prime)}_1\to f^{(\prime)}_1+\delta f^{(\prime)}_1$ of $\;\Xi^{+}_{c}\to \Xi^{0}\ell^+\nu_{\ell}\;$ and $\;\Xi^{0}_{c}\to \Xi^{-}\ell^+\nu_{\ell}$, we can directly fit these parameters from the data.  In doing the combination of $a^{\lambda,\lambda_w}_1 +a^{\lambda,\lambda_w}_5$ together to fit data, the $a^{\lambda,\lambda_w}_1$ in $\Lambda_c^+ \to n \ell^+ \nu_{\ell}$,  $\;\Xi^{+}_{c}\to \Sigma^{0}\ell^+\nu_{\ell}\;$ and $\;\Xi^{0}_{c}\to \Sigma^{-}\ell^+\nu_{\ell}\;$ will need to be treated as $a^{\lambda,\lambda_w}_1 - a^{\lambda,\lambda_w}_5$. One can take $a_1^{\lambda,\lambda_w}+a_5^{\lambda,\lambda_w}$, $a_2^{\lambda,\lambda_w}- a_4^{\lambda,\lambda_w}$, $a_2^{\lambda,\lambda_w}$, $a_3^{\lambda,\lambda_w}$, and $a_5^{\lambda,\lambda_w}$ as independent parameters.
The fitted results with two forms to access the $q^2$ dependence in form factors,  pole model and constant,  are given in Table~\ref{SU3SBF} with a reasonable $\chi^2/d.o.f=1.6$ and $\chi^2/d.o.f=1.9$, respectively.
It suggests that the SU(3) symmetry breaking effects generated by the light quark masses can improve the fit. Results for $\delta f_1$ and $\delta f^\prime_1$ characterize  the size of SU(3) symmetry breaking. From  Table~\ref{SU3SBF}, one can find that SU(3) symmetry breaking effects to the differential decay width for the $\Xi_c^0\to\Xi^-e^+\nu_{e}$  can reach as much as 50\%, depending on the kinematics.  Compared to the constant fit,   the pole model fit results of form factors will be used in Table~\ref{SU3SBF} since a relatively smaller $\chi^2$ is obtained. The inadequacy of the experimental data at this stage prevents  a direct analysis of different individual terms especially the 
 $a_2^{\lambda,\lambda_w}$, $a_3^{\lambda,\lambda_w}$, and $a_5^{\lambda,\lambda_w}$.   We hope that more experimental  data can be accumulated to further examine the detailed sources of  SU(3) symmetry breaking in the future.

%%%%%%%%%%%%%%%%%%%%%%%%%%%%%%%
\begin{table}[htbp!]
\caption{Experimental data and fit results of anti-triplet charmed baryons decays with symmetry breaking term. The form factors $f_1$ and $f_1^\prime$  correspond to $a^{\lambda,\lambda_w}_1 + a^{\lambda,\lambda_w}_5$. The form factors $\delta f_1$ and $\delta f_1^\prime$  correspond to $a^{\lambda,\lambda_w}_2 - a^{\lambda,\lambda_w}_4$. }\label{SU3SBF}
\begin{tabular}{|c|c|c|c|c|c|c|c|c|c|c|}\hline\hline
\multirow{2}{*}{channel} &\multicolumn{3}{c|}{ branching ratio$(\%)$} \cr\cline{2-4}
&experimental data& fit data(pole model)& fit data(constant).\\\hline
$\Lambda^{+}_{c}\to \Lambda^{0}e^+\nu_{e} $ & $3.6\pm0.4$&$3.61\pm0.32$&$3.62\pm0.32$\\\hline
$\Lambda^{+}_{c}\to \Lambda^{0}\mu^+\nu_{\mu} $ & $3.5\pm0.5$&$3.48\pm0.30$&$3.45\pm0.30$\\\hline
$\Xi^{+}_{c}\to \Xi^{0}e^+\nu_{e} $ & $2.3\pm1.5$&$3.89\pm0.73$&$3.92\pm0.73$\\\hline
$\Xi^{0}_{c}\to \Xi^{-}e^+\nu_{e} $ & $ 1.54\pm0.35$&$1.29\pm0.24$&$1.31\pm0.24$\\\hline
$\Xi^{0}_{c}\to \Xi^{-}\mu^+\nu_{\mu} $ & $ 1.27\pm0.44$&$1.24\pm0.23$&$1.24\pm0.23$\\\hline
fit parameter&\multicolumn{2}{c|}{$f_1=1.01\pm0.87$, $\delta f_1=-0.51\pm0.92$}&\multirow{2}{*}{$\chi^2/d.o.f=1.6$}\cr
(pole model)&\multicolumn{2}{c|}{$f^\prime_1=0.60\pm0.49$, $\delta f^\prime_1=-0.23\pm0.41$}&\\\hline
fit parameter&\multicolumn{2}{c|}{$f_1=0.86\pm0.92$, $\delta f_1=-0.25\pm0.88$}&\multirow{2}{*}{$\chi^2/d.o.f=1.9$}\cr
(constant)&\multicolumn{2}{c|}{$f^\prime_1=0.85\pm0.36$, $\delta f^\prime_1=-0.43\pm0.50$}&\\\hline
\end{tabular}
\end{table}
%%%%%%%%%%%%%%%%%%%%%%%%%%%%%%%%%%%%%%%%%%%%%%%

%%%%%%%%%%%%%%%%%%%%%%%%%%%%%%%%%%%%%
\begin{figure}[htbp!]
  \centering
  \includegraphics[width=0.6\columnwidth]{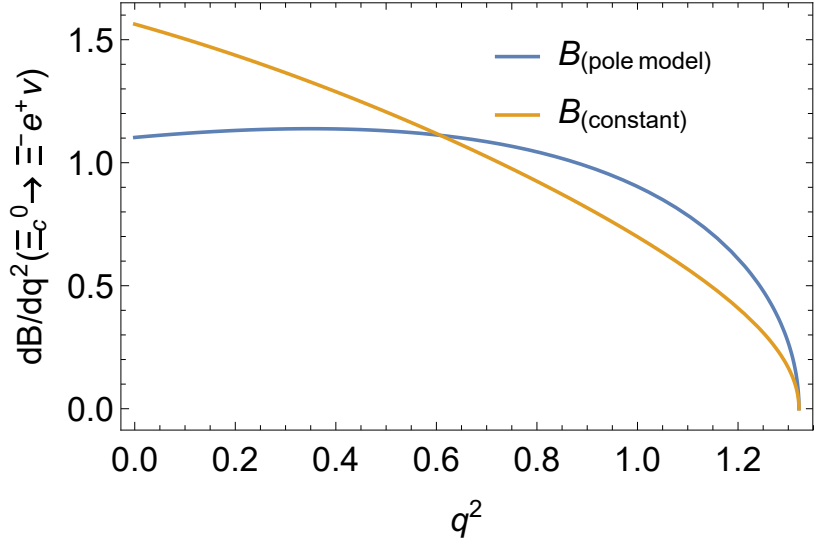}
\caption{ The differential decay branching fraction $d{\cal B}/dq^2$ for the $\Xi_c^0\to\Xi^-e^+\nu_{e}$. The two curves are obtained by different treatments of form factors. }
\label{decay width}
\end{figure}
%%%%%%%%%%%

In Fig~\ref{decay width},  we plot $d{\cal B}/dq^2$ in $\Xi_c^0\to\Xi^-e^+\nu_{e}$ with SU(3) symmetry form factors $f_i^{(')}$ as constants or parametrized as in Eq.~\eqref{eq:pole} to access the $q^2$ distribution. In both cases, the parameters in form factors are independently fitted. This can be tested by an  experimental  investigation in the future in addition to the branching ratio fitting in Table V.

%{\color{red} The area cover by each curve should have the same mass. But why the figures do not show so?}
%%%%%%%%%%%%%%%%%%%%%%%%%%%%%%%%%%%%%%%%%%%%%%%%

%{\color{red} Do not understand what you are trying to present for the following discussions.***}
%\\

 \subsection{Symmetry breaking caused by the $\Xi_{c}^{0/+}$--$\Xi_{c}^{\prime~0/+}$ mixing}

The inclusion of SU(3) breaking effects will lead to  $\Xi_{c}^{0/+}$ and $\Xi_{c}^{\prime~0/+}$ to mix. Here $\Xi_{c}^{\prime~0/+}$ is a component field  in the sextet $T_{c6}$,

\begin{eqnarray}
T_{c6} = 
\begin{pmatrix}
\Sigma_c^{++}& \frac{\Sigma_c^+}{\sqrt{2}} &\frac{\Xi_c^{+\prime}}{\sqrt{2}} \\
\frac{\Sigma_c^+}{\sqrt{2}} & \Sigma_c^0 & \frac{\Xi_c^{0\prime}}{\sqrt{2}}\\
\frac{\Xi_c^{+\prime}}{\sqrt{2}} & \frac{\Xi_c^{0\prime}}{\sqrt{2}} &\Omega_c^0
\end{pmatrix}.
\end{eqnarray}

The mixing between anti-triplet charmed baryons to the sextet states is due to the following term expanding to the first order in $m_s$, 
\begin{eqnarray}
H_{\lambda,\lambda_w}(T_{c\bar{3}}\omega\to T_{c6})=d^{\lambda,\lambda_w}\times(T_{c\bar3})^{[ij]}\omega_j^k (T_{c6})_{\{ki\}}. \label{eq:Hmix6}
\end{eqnarray}
Expanding Eq.~\eqref{eq:Hmix6}, one can find the  mixing between $\Xi_c^{0(+)}$ and $\Xi_c^{\prime 0(+)}$, while other hadrons are not affected. The mixing angle $\theta$ can be introduced to define the mass eigenvalue state $\Xi_c^0$ and $\Xi_c^+$,
\begin{eqnarray}
\Xi_c^{0/+ mass}=\cos\theta\times\Xi_c^{0/+}+\sin\theta\times\Xi_c^{0/+\prime},\label{eq:massmix}
\end{eqnarray}
where the angle $\theta$ is at the order $O(m_s)$. To the first order in $m_s$, $\cos\theta\sim 1$, $\sin\theta\sim \theta$. 

To take into account the mixing effects for physical $\Xi_c$ states, one needs to work out the sextet semileptonic decay amplitudes which are given to the first order in $\omega$
\begin{eqnarray}
H_{\lambda,\lambda_W}=&&c_1^{\lambda,\lambda_w} \times(T_{c6})^{\{ij\}}(H_3)^k \epsilon_{ikm}(T_8)^m_j+c_2^{\lambda,\lambda_w} \times(T_{c6})^{\{in\}}(H_3)^k \epsilon_{ikm}(T_8)^m_j \omega_n^j\notag\\&+&c_3^{\lambda,\lambda_w} \times(T_{c6})^{\{in\}}(H_3)^k \epsilon_{kjm}(T_8)^m_i \omega_n^j+c_4^{\lambda,\lambda_w} \times(T_{c6})^{\{in\}}(H_3)^k \epsilon_{jim}(T_8)^m_k \omega_n^j\notag\\
&+&c_5^{\lambda,\lambda_w} \times(T_{c6})^{[ij]}(H_3)^k \epsilon_{inm}(T_8)^m_j \omega_k^n.\label{eq:charmsb6}
\end{eqnarray}
% The relations between the helicity amplitude of different channels of sextet charmed baryons are presented in Table~\ref{tab:T6}.

The helicity amplitude for anti-triplet charmed baryon with mass eigenvalue state $\Xi^{mass}$ can be obtained by using Eq.~\eqref{eq:charmsb}, Eq.~\eqref{eq:massmix} and Eq.~\eqref{eq:charmsb6}. At the leading order,  the helicity amplitudes for the decay channel of mass eigenvalue states $\Xi_c^{0 mass}\to \Xi^-\ell^+\nu_{\ell}$ and $\Xi_c^{+ mass}\to\Xi^0\ell^+\nu_{\ell}$ become
\begin{eqnarray}
H^{mass}_{\lambda,\lambda_w}\propto V_{cs}^*(a_{1}^{\lambda,\lambda_w}+a_{2}^{\lambda,\lambda_w}-a_{4}^{\lambda,\lambda_w}+a_{5}^{\lambda,\lambda_w}+\frac{c_{1}^{\lambda,\lambda_w}}{\sqrt{2}}\theta), \label{eq:HT68}
\end{eqnarray}
where we have neglected  the $O(m_s^2)$ and higher order corrections.  
 The helicity amplitudes of other channels are listed in the `` amplitude I " column of Table~\ref{SU3phy}. In the table, the states in the first column are understood to be the mass eigenstates for the case with the mixing effect.

%%%%%%%%%%%%%%%%%%%%%%%%%%%%%%%%%%%%%%%%%%%%%%%%%%%%%%%%%%%%%%%%%
%%%%%%%%%%%%%%%%%%%%%%%%%%%%%%%%%%%%%%%%%%%%%%%%%%%%%%%%%%%%%%%%%

It is clear that the existing  experimental data is insufficient to determine all these parameters. But one can see that by introducing  the effective amplitude $a_{4}^{\prime\lambda,\lambda_w}=a_{4}^{\lambda,\lambda_w}+c_{1}^{\lambda,\lambda_w}\theta/\sqrt{2}$ and $a_{2}^{\prime\lambda,\lambda_w}=a_{2}^{\lambda,\lambda_w}+\sqrt{2}{c_{1}^{\lambda,\lambda_w}\theta}$, the effect of $\theta$ and $c_{1}^{\lambda,\lambda_w}$ can be absorbed into $a_{2}^{\prime\lambda,\lambda_w}$ and $a_{4}^{\prime\lambda,\lambda_w}$. The helicity amplitudes with $a_{2}^{\prime\lambda,\lambda_w}$, $a_{4}^{\prime\lambda,\lambda_w}$ are listed in the `` amplitude II " column of Table~\ref{SU3phy}.  Therefore, our fit results for the case without mixing effects are still valid, but the form factors $\delta f_1$ and $\delta f^\prime_1$ correspond to the new effective amplitudes  $a_{2}^{\prime\lambda,\lambda_w}$, $a_{4}^{\prime\lambda,\lambda_w}$.  

Although  several other form factors such as $\Delta f_1$ and $\Delta f^\prime_1$ cannot be constrained due to the lack of experimental data, in some scenarios, we still estimate the branching ratios of some processes from the results in Table~\ref{SU3phy}.   We can estimate the branching fractions of $\Lambda^{+}_{c}\to ne^+\nu_{e} $ and $\Lambda^{+}_{c}\to n\mu^+\nu_{\mu} $ :
\begin{eqnarray}
{\cal B}(\Lambda^{+}_{c}\to ne^+\nu_{e})&=&(0.520\pm0.046)\%,\quad {\cal B}(\Lambda^{+}_{c}\to n\mu^+\nu_{\mu})=(0.506\pm0.045)\%\;,
\end{eqnarray}
by assuming $a_5^{\lambda,\lambda_w}$ giving no contribution.  If the process of $\Lambda^{+}_{c}\to n \ell^+\nu_{\ell}$ is measured by experiments, the  contributions of $a_5^{\lambda,\lambda_w}$ will be obtained. The branching fractions of $\Xi^+_c\to \Sigma^0 \ell^+\nu_{\ell}$, $\Xi^+_c\to\Lambda^0 \ell^+\nu_{\ell}$ and $\Xi^0_c\to \Sigma^- \ell^+\nu_{\ell}$ can also be estimated by assuming $a_2^{\prime\lambda,\lambda_w}$, $a_3^{\lambda,\lambda_w}$, and $a_5^{\lambda,\lambda_w}$ giving no contributions. %comparing to the SU(3) symmetry parameter $a_1^{\lambda,\lambda_w}$.
\begin{eqnarray}
{\cal B}(\Xi^{+}_{c}\to \Sigma^0e^+\nu_{e})&=&(0.496\pm0.046)\%,\quad {\cal B}(\Xi^{+}_{c}\to \Sigma^0 \mu^+\nu_{\mu})=(0.481\pm0.044)\%,\notag\\
{\cal B}(\Xi^{+}_{c}\to \Lambda^0e^+\nu_{e})&=&(0.067\pm0.013)\%,\quad {\cal B}(\Xi^{+}_{c}\to \Lambda^0 \mu^+\nu_{\mu})=(0.069\pm0.0213)\%,\notag\\
{\cal B}(\Xi^{0}_{c}\to \Sigma^-e^+\nu_{e})&=&(0.333\pm0.031)\%,\quad {\cal B}(\Xi^{0}_{c}\to \Sigma^- \mu^+\nu_{\mu})=(0.323\pm0.029)\%.
\end{eqnarray}
 For the processes of $\Xi_{c}^{+}\rightarrow \Sigma^{0}\ell^{+}\nu_{\ell}$, $\Xi^{+}_{c}\rightarrow \Lambda^{0}\ell^{+}\nu_{\ell}$, $\Xi_{c}^{0}\rightarrow \Sigma^{-}\ell^{+}\nu_{\ell}$, once some of the processes are established in future experiments,  we can fit the form factor $\Delta f_1^{(\prime)}$ which reflects the contribution of $a_3^{\lambda,\lambda_w}$. Then the branching fractions for the other processes depending on $a_3^{\lambda,\lambda_w}$ can also be established.

\section{SU(3) symmetry analysis in anti-triplet beauty baryons semileptonic decays}

The anti-triplet beauty baryon semileptonic decays are governed  by the Hamiltonian:
\begin{eqnarray}
{\cal H}_{b\to u/c}&=&\frac{G_F}{\sqrt2} \left[V_{qb}^{*} \bar q  \gamma^\mu(1-\gamma_5)b ~\bar \ell\gamma_\mu(1-\gamma_5)\nu_{\ell} \right] +h.c.,
\end{eqnarray}
where $q=u,c$.

The $b\to c$ transition is an SU(3) singlet, while the $b\to u$ transition forms an SU(3) triplet $H^\prime_3$ with $(H^\prime_3)^1=1$ and $(H^\prime_3)^{2,3}=0$.
The SU(3) matrix representation of anti-triplet beauty baryons are given as
\begin{eqnarray}\
T_{b\bar3}=
\begin{pmatrix}
0& \Lambda_b^0 &\Xi_b^0 \\
-\Lambda_b^0 & 0&\Xi_b^-\\
-\Xi_b^0& -\Xi_b^-&0
\end{pmatrix}\;.
\end{eqnarray}
We write the helicity amplitude in SU(3) analysis in a similar fashion as what has been done for semileptonic charmed anti-triplet decays,  as
\begin{eqnarray}
H_{\lambda,\lambda_w}=b_1^{\lambda,\lambda_w} \times(T_{b\bar3})^{[ij]}(H_3^\prime)^k \epsilon_{ikm}(T_8)^m_j+e_1^{\lambda,\lambda_w} \times(T_{b\bar3})^{[ij]}(T_{c\bar 3})_{[ij]},
\end{eqnarray}
where $b_1^{\lambda,\lambda_w}$ and $e_1^{\lambda,\lambda_w}$ are respectively similar to $a_1^{\lambda,\lambda_w} $ in the previous section.
The Feynman diagrams for the two term in $H_{\lambda,\lambda_w}$ are shown in (a) and (b) of Fig~\ref{feynman} respectively.

Expanding the $H_{\lambda,\lambda_w}$,  one can obtain SU(3) amplitudes are listed in Table~\ref{SU3b} and the SU(3) relations can be given as follows,
\begin{eqnarray}
\Gamma(\Lambda^{0}_{b}\to p\ell^-\bar{\nu_{\ell} } )&=&\Gamma(\Xi^{0}_{b}\to \Sigma^{+}\ell^-\bar{\nu_{\ell} })=2\Gamma(\Xi^{-}_{b}\to \Sigma^{0}\ell^-\bar{\nu_{\ell} } )=6\Gamma(\Xi^{-}_{b}\to \Lambda^{0}\ell^-\bar{\nu_{\ell} })\;,\notag\\
\Gamma(\Lambda^{0}_{b}\to \Lambda^{+}_{c}\ell^-\bar{\nu_{\ell} } )&=&\Gamma(\Xi^{0}_{b}\to \Xi^{+}_{c}\ell^-\bar{\nu_{\ell} } )=\Gamma(\Xi^{-}_{b}\to \Xi^{0}_{c}\ell^-\bar{\nu_{\ell} })\;.\label{SU3RB}
\end{eqnarray}
Using the experimental data $\mathcal{B}$($\Lambda_b^0\to\Lambda_c^+\ell^-\bar{\nu_{\ell} }$)=$(6.2^{+1.4}_{-1.3})\%$ and $\mathcal{B}$($\Lambda_b^0\to p \mu^-\bar{\nu_{\mu}}$)=$(4.1\pm1.0)\%$, we give the prediction in third column of Table~\ref{SU3b}.

%%%%%%%%%%%%%%
\begin{table}[htbp!]
\caption{Amplitudes beauty baryons $\Xi_{b}$ and $\Lambda_{b}$ decays into octet and anti-triplet baryons.}\label{SU3b}
\begin{tabular}{|c|c|c|c|c|c|c|c}\hline\hline
channel & amplitude &  branching fraction ($\%$)\\\hline
$\Lambda^{0}_{b}\to p\ell^-\bar{\nu_{\ell} } $ & $ b_1^{\lambda,\lambda_w}$& $4.1\pm1.0$(input)\cite{Zyla:2020zbs}\\\hline
$\Xi^{0}_{b}\to \Sigma^{+}\ell^-\bar{\nu_{\ell} } $ & $ -b_1^{\lambda,\lambda_w}$& $ 4.1\pm1.0$\\\hline
$\Xi^{-}_{b}\to \Sigma^{0}\ell^-\bar{\nu_{\ell} } $ & $ \frac{b_1^{\lambda,\lambda_w}}{\sqrt{2}}$& $ 2.2\pm0.5$\\\hline
$\Xi^{-}_{b}\to \Lambda^{0}\ell^-\bar{\nu_{\ell} } $ & $ \frac{b_1^{\lambda,\lambda_w}}{\sqrt{6}}$& $ 0.7\pm0.2$\\\hline
\hline$\Lambda^{0}_{b}\to \Lambda^{+}_{c}\ell^-\bar{\nu_{\ell} } $ & $ 2 e_1^{\lambda,\lambda_w}$ & $6.2^{+1.4}_{-1.3}$(input)\cite{Zyla:2020zbs}\\\hline
$\Xi^{0}_{b}\to \Xi^{+}_{c}\ell^-\bar{\nu_{\ell} } $ & $ 2 e_1^{\lambda,\lambda_w}$& $6.2^{+1.4}_{-1.3}$\\\hline
$\Xi^{-}_{b}\to \Xi^{0}_{c}\ell^-\bar{\nu_{\ell} } $ & $ 2 e_1^{\lambda,\lambda_w}$& $6.6^{+1.5}_{-1.4}$\\\hline
\hline
\end{tabular}
\end{table}
%%%%%%%%%%%%%%%%%%%%%%%%%%%%%%%%%%%%%%%%%%%%%%%%%%%

For the processes we predicted, we expect them to be measured by Belle~II and LHCb. The SU(3) symmetry of these processes will probably be tested. Due to the lack of experimental data at this stage,  we can not explore the SU(3) symmetry breaking effects by fitting the form factors. We have also worked out how to include SU(3) symmetry breaking effects. The helicity amplitude including SU(3) symmetry breaking about b quark decays is given as:
\begin{eqnarray}
H_{\lambda,\lambda_w}=&&b_1^{\lambda,\lambda_w} \times(T_{b\bar3})^{[ij]}(H^\prime_3)^k \epsilon_{ikm}(T_8)^m_j+b_2^{\lambda,\lambda_w} \times(T_{b\bar3})^{[in]}(H^\prime_3)^k \epsilon_{ikm}(T_8)^m_j \omega_n^j\notag\\&+&b_3^{\lambda,\lambda_w} \times(T_{c\bar3})^{[in]}(H^\prime_3)^k \epsilon_{jkm}(T_8)^m_i \omega_n^j+b_4^{\lambda,\lambda_w} \times(T_{c\bar3})^{[in]}(H^\prime_3)^k \epsilon_{ikm}(T_8)^m_j \omega_n^j\notag\\
&+&b_5^{\lambda,\lambda_w} \times(T_{c\bar{3}})^{[ij]}(H_3^\prime)^k \epsilon_{inm}(T_8)^m_j\omega_k^n\notag\\
&+&e_1^{\lambda,\lambda_w} \times(T_{b\bar3})^{[ij]}(T_{c\bar 3})_{[ij]}+e_2^{\lambda,\lambda_w}\times (T_{b\bar3})^{[ij]}(T_{c\bar 3})_{[kj]}\omega^k_i\;,
\end{eqnarray}
 where the $b_1^{\lambda,\lambda_w}$,$e_1^{\lambda,\lambda_w}$ are SU(3) symmetry irreducible nonperturbative amplitude and $b_2^{\lambda,\lambda_w}$,$b_3^{\lambda,\lambda_w}$,$b_4^{\lambda,\lambda_w}$,$b_5^{\lambda,\lambda_w}$, $e_2^{\lambda,\lambda_w}$ are SU(3) symmetry breaking irreducible nonperturbative amplitudes. Here we have written the $b_i^{\lambda,\lambda_w}$ terms in a similar fashion as that for $a_i^{\lambda,\lambda_w}$ terms. But the $b_5^{\lambda,\lambda_w}$ term has no contribution, because $(H_3^\prime)^{ k}\omega_k^n$ is equal to zero.
Expanding the formula above, we collected the SU(3) amplitudes  in Table~\ref{tab:bcq_semi}.

%%%%%%%%%%%%%%%%%%%
\begin{table}[htbp!]
\caption{SU(3) symmetry breaking amplitudes of beauty baryons $\Xi_{b}$ and $\Lambda_{b}$ decay into octet baryons and anti-triplet charmed baryons, respectively.}\label{tab:bcq_semi}
\begin{tabular}{|c|c|c|c|c|c|c|c}\hline\hline
channel & amplitude \\\hline
$\Lambda^{0}_{b}\to p\ell^-\bar{\nu_{\ell} } $ & $ b_1^{\lambda,\lambda_w}$\\\hline
$\Xi^{0}_{b}\to \Sigma^{+}\ell^-\bar{\nu_{\ell} } $ & $ -b_1^{\lambda,\lambda_w}+b_3^{\lambda,\lambda_w}-b_4^{\lambda,\lambda_w}$\\\hline
$\Xi^{-}_{b}\to \Sigma^{0}\ell^-\bar{\nu_{\ell} } $ & $ \frac{b_1^{\lambda,\lambda_w}-b_3^{\lambda,\lambda_w}-b_4^{\lambda,\lambda_w}}{\sqrt{2}}$\\\hline
$\Xi^{-}_{b}\to \Lambda^{0}\ell^-\bar{\nu_{\ell} } $ & $ \frac{b_1^{\lambda,\lambda_w}+2 b_2^{\lambda,\lambda_w}+b_3^{\lambda,\lambda_w}-b_4^{\lambda,\lambda_w}}{\sqrt{6}}$\\\hline
\hline$\Lambda^{0}_{b}\to \Lambda^{+}_{c}\ell^-\bar{\nu_{\ell} } $ & $ 2 e_1^{\lambda,\lambda_w}$\\\hline
$\Xi^{0}_{b}\to \Xi^{+}_{c}\ell^-\bar{\nu_{\ell} } $ & $ 2 e_1^{\lambda,\lambda_w}+e_2^{\lambda,\lambda_w}$\\\hline
$\Xi^{-}_{b}\to \Xi^{0}_{c}\ell^-\bar{\nu_{\ell} } $ & $ 2 e_1^{\lambda,\lambda_w}+e_2^{\lambda,\lambda_w}$\\\hline
\hline
\end{tabular}
\end{table}
%%%%%%%%%%%%%%%%%%%
A number of relations for decay widths can be readily deduced from Table~\ref{tab:bcq_semi},
\begin{eqnarray}
\Gamma(\Xi_b^-\to\Sigma^{0}\ell^-\bar{\nu_{\ell} }  )&=&\frac{1}{2}\Gamma(\Xi^{0}_{b}\to \Sigma^{+}\ell^-\bar{\nu_{\ell} })\; ,\notag\\
\Gamma(\Xi^{0}_{b}\to \Xi^{+}_{c}\ell^-\bar{\nu_{\ell} } )&=&\Gamma(\Xi^{-}_{b}\to \Xi^{0}_{c}\ell^-\bar{\nu_{\ell} })\label{SU3SBR}\; .
\end{eqnarray}

It can be seen from Eq.(\ref{SU3SBR}) that though the SU(3) symmetry breaking effects caused by the light quark mass are taken into account,  there are still relations in these processes. These relations result from isospin symmetry which can only be broken if non-zero u and d quark masses with different values are included. We strongly suggest our experimental colleagues carry out measurements for these decays.

\section{Conclusion}
We have investigated the semileptonic decay of anti-triplet heavy baryons using SU(3) symmetry based on the latest experimental data. In the SU(3) symmetry limit,  when fitting the available experimental data to the SU(3) symmetry analysis, we can only obtain a fit with a least $\chi^2/d.o.f = 14.3$ which means SU(3) symmetry is not a good symmetry for semileptonic charmed anti-baryon decays. We have then carried out detailed  analyses with SU(3) symmetry breaking effect due to mass difference between $s$ quark and $u/d$ quark mass.  In one scenario, we obtain a reasonable description of all relevant data with a $\chi^2/d.o.f = 1.6$.  As an estimation, we give the branching ratios for $\Lambda^{+}_{c}\rightarrow n \ell^+\nu_{\ell}\;$, $\Xi^+_c\to \Sigma^0 \ell^+\nu_{\ell}\;$, $\Xi^+_c\to\Lambda^0 \ell^+\nu_{\ell}\;$ and $\Xi^0_c\to \Sigma^- \ell^+\nu_{\ell}\;$ in some scenarios.

We have also extended the analysis to the semileptonic decays of anti-triplet beauty baryons. However, the lack of experimental data prevents us from an in-depth study. Instead, we find a set of SU(3) relations in Eq.~(\ref{SU3RB}) and isospin relation in Eq.~(\ref{SU3SBR})  between the decay widths of such processes. Our results will help to explore the physics behind these SU(3) symmetry breaking experimental data with more experimental data available in the future.

%%%%%%%%%%%%%%%%%%%%%%%%%%%%%%%%%%%%%%%%%%%%%
\section*{ACKNOWLEDGMENT}
F.Huang and Z.P.Xing are grateful to J.Sun and Z.X.Zhao for their valuable comments. This work was supported in part by NSFC under Grant Nos. 11735010, U2032102, 11975149, 12090064, 12125503, and  NSFS under Grant No. 15DZ2272100 and  the MOST (Grant No. MOST 106-2112-M-002-003-MY3 ).

%%%%%%%%%%%%%%%%%%%%%%%%%%%%%%%%%%%%%%%%

\end{document}